\documentclass[a4paper, 11pt]{article}
\usepackage[english]{babel}
\usepackage[utf8]{inputenc}
\usepackage[capposition=top]{floatrow}

\usepackage{amsthm}
\usepackage{amsmath}
\usepackage{amssymb}
\usepackage{graphicx}
\usepackage{ragged2e}
\graphicspath{./immages/}
\usepackage{chngcntr}
\usepackage[colorinlistoftodos]{todonotes}
\usepackage{changepage}
\usepackage{tabularx}
\usepackage{makecell}
\usepackage{array}
\usepackage{multirow}
\usepackage{changepage}
\usepackage{ctable}
\usepackage{subcaption}
\usepackage{threeparttable}
\usepackage{longtable}
\usepackage{pdflscape}
\usepackage{geometry}
\usepackage{fancyhdr}
\usepackage{everypage}
\usepackage{titlesec}
\usepackage{adjustbox}

\newcolumntype{L}[1]{>{\raggedright\arraybackslash}p{#1}}
\newcolumntype{C}[1]{>{\centering\arraybackslash}p{#1}}
\newcolumntype{R}[1]{>{\raggedleft\arraybackslash}p{#1}}

\newgeometry{landscape, left=1in, right=1in, top=1in, bottom=1in}

\newcommand{\Lpagenumber}{\ifdim\textwidth=\linewidth\else\bgroup
  \dimendef\margin=0 
  \ifodd\value{page}\margin=\oddsidemargin
  \else\margin=\evensidemargin
  \fi
  \raisebox{\dimexpr -\topmargin-\headheight-\headsep-0.5\linewidth}[0pt][0pt]{%
    \rlap{\hspace{\dimexpr \margin+\textheight+\footskip}%
    \llap{\rotatebox{90}{\thepage}}}}%
\egroup\fi}
\AddEverypageHook{\Lpagenumber}%

\newcolumntype{P}[1]{>{\centering\arraybackslash}p{#1}}
\newcolumntype{M}[1]{>{\centering\arraybackslash}m{#1}}

\usepackage{setspace}

\usepackage[authordate,sorting=author-chron,backend=biber]{biblatex-chicago}

\DeclareSortingScheme{author-chron}{
  \sort{
    \field{presort}
  }
  \sort{
    \field{sortname}
    \field{author}
    \field{editor}
    \field{translator}
  }
  \sort{
    \field{sortyear}
    \field{year}
  }
  \sort{
    \field{sortmonth}
    \field{month}
  }
  \sort{
    \field{sorttitle}
    \field{title}
  }
}
\bibliography{bib}

\DeclareFieldFormat{citehyperref}{%
  \DeclareFieldAlias{bibhyperref}{noformat}
  \bibhyperref{#1}}

\DeclareCiteCommand{\citeyearwithoutparen}
  {\usebibmacro{prenote}}
  {\usebibmacro{citeindex}%
   \printtext[citehyperref]{\printfield{labelyear}}}
  {\multicitedelim}
  {\usebibmacro{postnote}}

\DeclareCiteCommand{\citeauthorwithoutparen}
  {\boolfalse{citetracker}%
   \boolfalse{pagetracker}%
   \usebibmacro{prenote}}
  {\ifciteindex
     {\indexnames{labelname}}
     {}%
   \printnames{labelname}}
  {\multicitedelim}
  {\usebibmacro{postnote}}

\usepackage[colorlinks,bookmarksopen,bookmarksnumbered,citecolor=red,urlcolor=red]{hyperref}
\hypersetup{
    colorlinks=true,
    linkcolor=blue,
    filecolor=blue,
    urlcolor=blue,
    citecolor=blue,  
}

\title{US labor market conditions and migration: a reassessment of Bahar (2025)}
\date{\today}		
\author{Francisco Rodríguez\thanks{Corresponding author. Center for Economic and Policy Research and University of Denver. E-mail;francisco.rodriguez4@du.edu} and Giancarlo Bravo\thanks{Oil for Venezuela}
}
\begin{document}

\maketitle
\onehalfspacing


\begin{abstract}
Bahar (2025) argues that there is a long-term cointegrating relationship between US job vacancies and Southwest border crossings. We show that his conclusion is based on a misspecified Engle–Granger test applied to first differences. Once the Engle–Granger test is correctly applied to levels, evidence for a cointegrating relationship vanishes, invalidating the paper’s approach to estimating short- and long-run elasticities. Bahar’s approach is therefore uninformative about the relationship between US labor market conditions and migration.
\end{abstract}

\section{Introduction}

Using monthly data for the United States, \textcite{bahar2025,bahar2025d} argues that there is a strong positive correlation between Southwest U.S. border crossings and labor market tightness, leading to the conclusion that "pull" factors play a significant role in driving migration patterns towards the U.S. \parencite[679]{bahar2025}.  We show that 
this conclusion is premised on findings from a misspecified cointegration test.  Correcting this mistake makes the finding of a long-term relationship disappear and undermines the justification for the methods used in the paper to estimate both short and long-term elasticities.\footnote{ 
Note that \textcite{bahar2025d} is a corrected version of \textcite{bahar2025}, and that as of the time of this writing there are three different versions of replication code published for the paper.  Where necessary, we will distinguish between the original (or uncorrected) and corrected versions of the paper, and will refer to the various replication codes in chronological order of publication.} 

Section \ref{eg} explains why Bahar's choice to apply a cointegration test to first differences of I(1) variables is incorrect. Section \ref{results} shows how applying a correct cointegration test in levels fails to yield the conclusive evidence of cointegration claimed by Bahar.   Section \ref{repli} discusses how our results invalidate Bahar's econometric approach and findings.

\section{Methods}\label{eg}

Bahar's paper studies the relationship between flows of irregular migrants into the United States and U.S. labor market conditions.  His empirical analysis focuses on two indicators: encounters at the Southwest United States border  and job openings per unemployed person in the United States.  Bahar claims that both of these series are I(1), and that ``when the Engle-Granger cointegration test was applied to these two series, this produced a test statistic of -18.829, well above [sic] the critical value of $-4.024$ at the 1\% significance level.''\parencite[676-677]{bahar2025}.\footnote{Bahar provides no results from unit root tests---either in the paper, online appendix, or replication code---to support his claim that the series are I(1). Our application of standard unit root tests to his data gives conflicting results regarding whether these series can be characterized as I(1).}  Bahar's Appendix F makes a similar claim regarding the quarterly series for these variables. Neither of these claims was affected by the June 2025 correction of Bahar's paper.

Bahar's cointegration findings come from applying Engle-Granger cointegration tests to the first differences of the series.\footnote{We conclude that he applied the test to the first differences of the series in question based on several pieces of information. First, although neither his corrected replication code (version 3.0) nor the replication code published alongside the original paper (version 2.0) implement Engle-Granger tests, a version of the code that was briefly available prior to publication (version 1.0) does contain a misspecified Engle-Granger test in first differences.\footnote{The code in that version reads ``egranger d.openingsrate\_nsa d.encounters\_ nsa". The correct code to run a conventional Engle-Granger test in levels would be ``egranger openingsrate\_nsa encounters\_ nsa".}  Second, \textcite[p.3]{bahar2025b}, who claim to use the same methodology as \textcite{bahar2025}, explicitly state that they apply the Engle–Granger test to the first differences of the variables [p.3], as can also be verified in the replication code for that paper. Third, we are able to either nearly replicate or  exactly replicate Bahar’s cointegration results using a first-difference test, while applying the standard test in levels yields very different values. More concretely, applying the test in first differences from version 1.0 of the replication code to Bahar's full sample yields a test statistic of -19.208, while applying it to a shorter time window (which Bahar's uncorrected code used for some results) gives a test statistic of -18.834, very close to the -18.829 value that he reports. Applying the test in first differences to the shortened time window on the quarterly data exactly reproduces Bahar's test statistic of -8.656 reported in Appendix F. Note that Bahar misreports the 1\% critical values  as -4.024 in both cases (the correct values are -3.936 and -4.020, respectively).} The Engle-Granger cointegration test is designed to be run on the levels of variables that are I(1). If the series are I(1) -- and Bahar claims that they are \footnote{See \textcite[p. 676]{bahar2025}: ``Both series are integrated of order 1.''} -- then their first differences must be stationary, and the Engle-Granger test applied to them will trivially find evidence of cointegration. This is because linear combinations of stationary variables are, under general conditions, stationary, and the Engle-Granger test is a test for stationarity in a linear combination of the variables it is applied on (the OLS residuals from the first stage regression).  \textcite{rodriguez2025a} present Monte Carlo simulations showing that applying the Engle-Granger test to the first differences of two non-cointegrated unit root series results in false positive rates of 100 percent.

Applying the Engle-Granger test to first differences of I(1) series is therefore incorrect. The original presentation of the test by its creators, standard econometrics textbooks and statistical software manuals specify that the method is designed to be applied on the levels of the corresponding variables (see, e.g., \textcite[p. 268]{engle1987}, \textcite[p. 1001]{Greene2012}, \textcite{schaffer2010}).  An exhaustive literature search has found no literature supporting the use of the test on first-differenced data nor any implementations by paper others than \textcite{bahar2025} and \textcite{bahar2025a}. 
\begin{table}[ht]
\centering
\caption{Engle-Granger Test Statistics Across Specifications}
\label{tab1}
\resizebox{\textwidth}{!}{ 
\begin{tabular}{lrrrr}
\toprule
\textbf{} & \multicolumn{2}{c}{\textbf{Logarithms}} & \multicolumn{2}{c}{\textbf{Untransformed}} \\
\cmidrule(lr){2-3} \cmidrule(lr){4-5}
 & Encounters & Openings & Encounters & Openings \\
\midrule
Engle-Granger & -3.153* & -2.183 & -4.061*** & -3.156* \\
Augmented Engle-Granger, 12 lags & -2.614 & -1.922 & -2.182 & -1.851 \\
Augmented Engle-Granger, 12 lags and trend & -2.968 & -3.052 & -2.860 & -3.790* \\
\addlinespace
Bahar's reported Engle-Granger test statistic & \multicolumn{4}{c}{-18.829***} \\
\bottomrule
\end{tabular}
}
\begin{tablenotes}
\small
\item Results represent the Engle-Granger test statistic for various specifications. Columns 1 and 3 normalize the coefficient of the cointegrating vector on encounters to one, while Columns 2 and 4 normalize the coefficient on openings to one. The last row reports the result presented by \textcite[677]{bahar2025}. Significance levels: *** 1 percent, ** 5 percent, * 10 percent.

\end{tablenotes}
\end{table}

\section{Results}
\label{results}

Table \ref{tab1} reports the results of twelve specifications of Engle-Granger cointegration tests applied to Bahar's data and representing combinations of choice of variable (logarithms or untransformed), lag and trend controls, and normalization assumption. We report this range of specifications because the text of Bahar's paper is itself unclear regarding whether the test was run on the untransformed values of the variables or their logarithms, and which variable was normalized as the dependent variable.\footnote{Since the author uses an OLS estimate of the cointegrating vector in logarithms for his ARDL specification, it would be natural to conclude that the test had been carried out in logarithms; however, the miscoded version of the code in version 1.0 applies the test to the untransformed variables.} It is also customary to assess the sensitivity of the Engle-Granger test to the inclusion of lags and a trend term. The set of specifications reported in Table \ref{tab1} thus gives a reasonable sense of the range of results that would emerge from running conventional Engle-Granger tests on Bahar's data.

The range of test statistics obtained in these estimates spans from -1.851 to -4.061, with a median value of -2.914 --- less than one-sixth the magnitude of the -18.829 statistic reported by Bahar.  Of the twelve specifications, only one rejects the null of no cointegration at a significance level of 5\% or lower, in sharp contrast with Bahar's claim of having comfortably  rejected the null of no cointegration at a 1\% significance level.

\section{Discussion}
\label{repli}

Correcting the misspecified tests identified in this note severely undermines Bahar's empirical strategy and sheds doubt on the validity of his findings.  This is because the claim of cointegration between border encounters and job openings is central to the paper's empirical approach. 

Bahar’s empirical strategy is completely based on estimation of his equations (B1) and (B2).  Equation (B1) is a levels regression between two I(1) series which, in the absence of cointegration, is well-known to be prone to spurious estimates \parencite[p.557]{hamilton1994}.  Equation (B2) is an autoregressive distributed lag specification which includes an error-correction term (ECT) that is only defined if there is a cointegrating relationship between the variables.  In the absence of such a relationship, the estimate of the ECT (i.e., the residual from the levels regression) will be I(1) and its inclusion in the regression will risk producing spurious results.  Therefore, both the estimates of short- and long-run elasticities presented by Bahar are derived from misspecified regressions and cannot be relied upon. 

Bahar's claim that there is a positive correlation between labor market tightness and the frequency of border crossings is an interesting hypothesis worthy of systematic investigation.  Given that the methods applied in his paper to reach that conclusion are based on misspecified regressions premised on a cointegrating relationship for which there is no evidence, Bahar's article is uninformative on what the evidence tells us about this key question.

\textbf{Acknowledgement}

This manuscript was prepared with the assistance of ChatGPT (versions 4 and 5) for the purpose of exploration of ideas, language improvement and coding assistance.

\textbf{Declaration of interests}

The authors declare no relevant  competing interests.

\textbf{Declaration of funding} 

This research received no external funding.

\textbf{Data availability statement} 

A replication package is available at:

https://dataverse.harvard.edu/dataset.xhtml?persistentId=doi:10.7910/DVN/ZNTAVQ.

\printbibliography

\end{document}